\documentclass[10pt,prd,tightenlines,twocolumn,superscriptaddress]{revtex4}

\usepackage{amsmath,amssymb}

\usepackage{color}
\usepackage{graphicx}
\definecolor{darkblue}{rgb}{0,0,0.7}
\definecolor{darkred}{rgb}{0.7,0,0}
\definecolor{darkgreen}{rgb}{0,0.5,0}
\definecolor{added}{rgb}{0.8,0.4,0}
\usepackage[unicode, colorlinks, citecolor=darkblue, linkcolor=darkred, urlcolor=blue]{hyperref}

\allowdisplaybreaks

\begin{document}

\title{Stable optical spring in  aLIGO detector with unbalanced arms and in Michelson-Sagnac interferometer}

\author{Nikita Vostrosablin \footnote{Electronic address: vostrosablin@physics.msu.ru}}
\affiliation{Physics Department, Moscow State University, Moscow 119992 Russia}

\author{Sergey P. Vyatchanin}
\affiliation{Physics Department, Moscow State University, Moscow 119992 Russia}

\date{ \today}

\begin{abstract}


Optical rigidity in aLIGO gravitational-wave detector, operated on dark port regime, is unstable. We show that the same interferometer with excluded symmetric mechanical mode but with unbalanced arms allows to get stable optical spring for antisymmetric mechanical mode. Arm detuning necessary to get stability is shown to be a small one --- it corresponds to small power in signal port. We show that stable optical spring may be also obtained in Michelson-Sagnac interferometer with both power and signal recycling mirrors and unbalanced arms. 

\end{abstract}

\maketitle

\section{Introduction}

 Ground-based gravitational waves antennas form worldwide net of  large-scale detectors like LIGO~\cite{Abbott_2009,Harry_2010}, 
VIRGO~\cite{Accadia2012} and GEO~\cite{Grote2010}. Extremely high sensitivity of this detectors is limited by noises of 
different nature.  In the low frequency range  (around $10$ Hz) the gravity-gradient (Newtonian) noise prevails, below $\sim 50$ Hz --- seismic ones, at middle frequencies ($\sim 50 - 200$ 
Hz) thermal noises dominate and in high frequency range (over $200$ Hz) photon shot noise makes main contribution. Next 
generation of gravitational wave antennas (Advanced LIGO or aLIGO \cite{Harry_2010}, Advanced VIRGO \cite{advVIRGO})  and also third generation detectors (such as Einstein Telescope \cite{ET1, ET2}, GEO-HF \cite{GEO_HF} and KAGRA \cite{KAGRA}) promise by 
compensation and suppression of thermal and other noises  to achieve sensitivity of Standard Quantum Limit (SQL) 
\cite{1968_SQL,1975_SQL,1977_SQL,1992_quant_meas} for continuous measurement defined only by quantum noise.
SQL is the optimal combination of two noises of quantum nature: fluctuations of light pressure caused by random photon 
number falling onto mirror's surface and photon counting noise.

Possible way to overcome the SQL is the usage of optical rigidity (optical spring 
effect)~\cite{64a1BrMiVMU,67a1BrMaJETP,78BrBook,1992_quant_meas}. Recall optical rigidity appears in a detuned Fabry-Perot 
interferometer --- the circulating power and consequently the radiation pressure became dependent on the distance between the 
mirrors. It has been shown~\cite{99a1BrKhPLA,01a1KhPLA,01ChenPRD,02ChenPRD,05a1LaVyPLA,06a1KhLaVyPRD,2007_NiSaKa_PRD, M_Rakh} that 
gravitational wave detectors using optical springs exhibit sensitivity below the SQL.

In case of single pump an interferometer utilizing optical rigidity has two subsystems: a mechanical one and 
an optical one. Interaction between them gives birth to two eigen modes each of which is characterized by its own resonance 
frequency and damping. For description of evolution one can make transfer from the conventional coordinates to eigen ones 
and consider the evolution of the system as evolution of these (normal) oscillators~\cite{12a1RaVyPLA}.

Dynamics of complex system such as aLIGO detector can be considered on the basis of more simpler and well studied system --- Fabry-Pero resonator. Such equivalence is termed {\it scaling law}       \cite{scal}.
Fabry-Pero resonator with only one optical spring is always unstable because a single pump introduces either positive spring with negative damping or negative spring with positive damping~\cite{64a1BrMiVMU, 67a1BrMaJETP, 78BrBook,99a1BrKhPLA}. The obvious ways to avoid instabilities is 
implementation of feedback~\cite{02ChenPRD}. Another way is utilization of additional pump \cite{08ChenPRD,2011_RaHiVy_PRD}, 
which has been investigated in details and proven experimentally with mirror of gram-scale~\cite{07CorbitPRL}.

\begin{figure}[h]
	\begin{center}
		 \includegraphics[width = .85\linewidth]{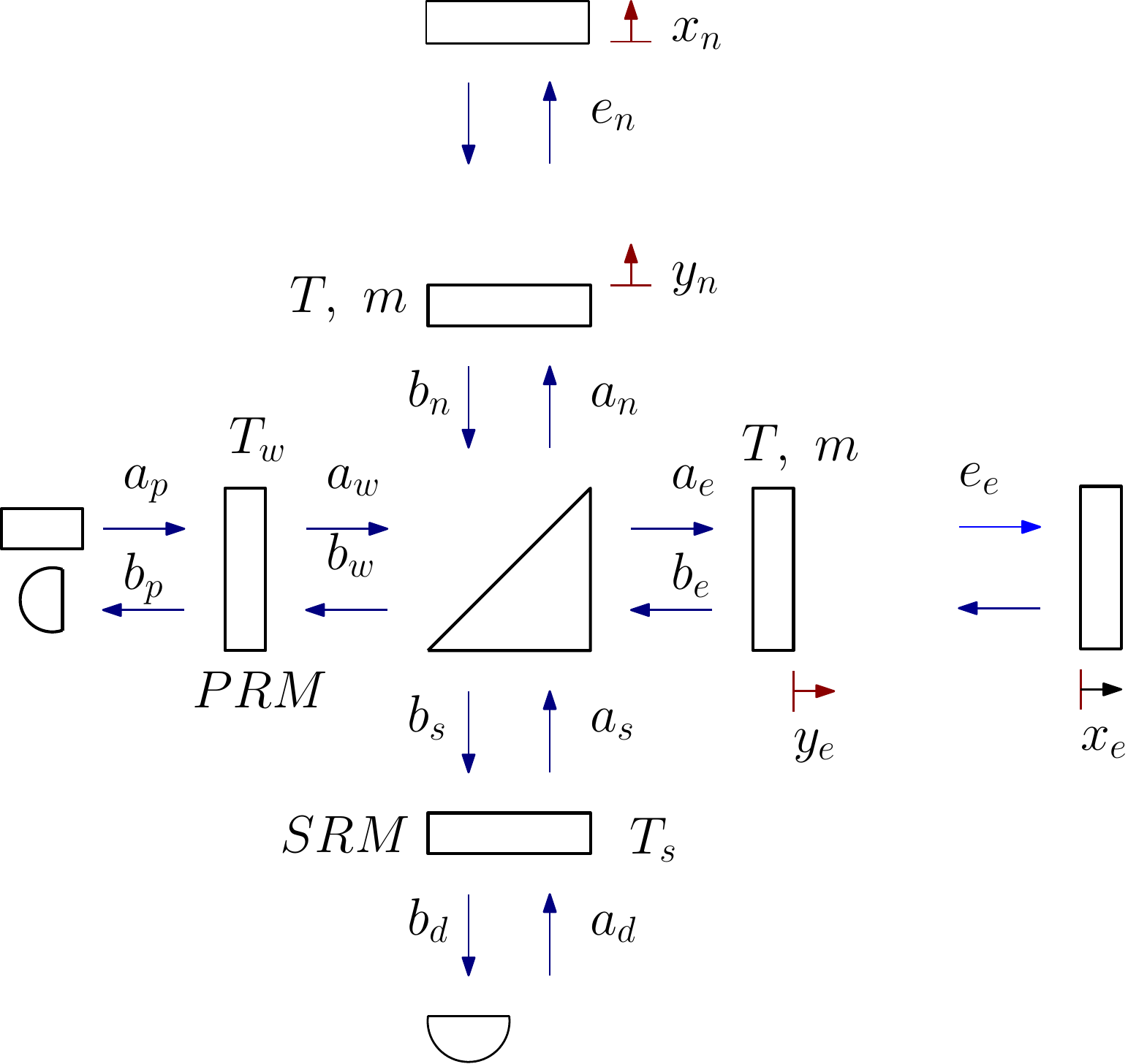}
		\caption{
		Scheme of Advanced LIGO detector. PRM  (SRM) are power (signal) recycling mirror.
		}\label{aligo}
		
		\includegraphics[width=0.35\textwidth]{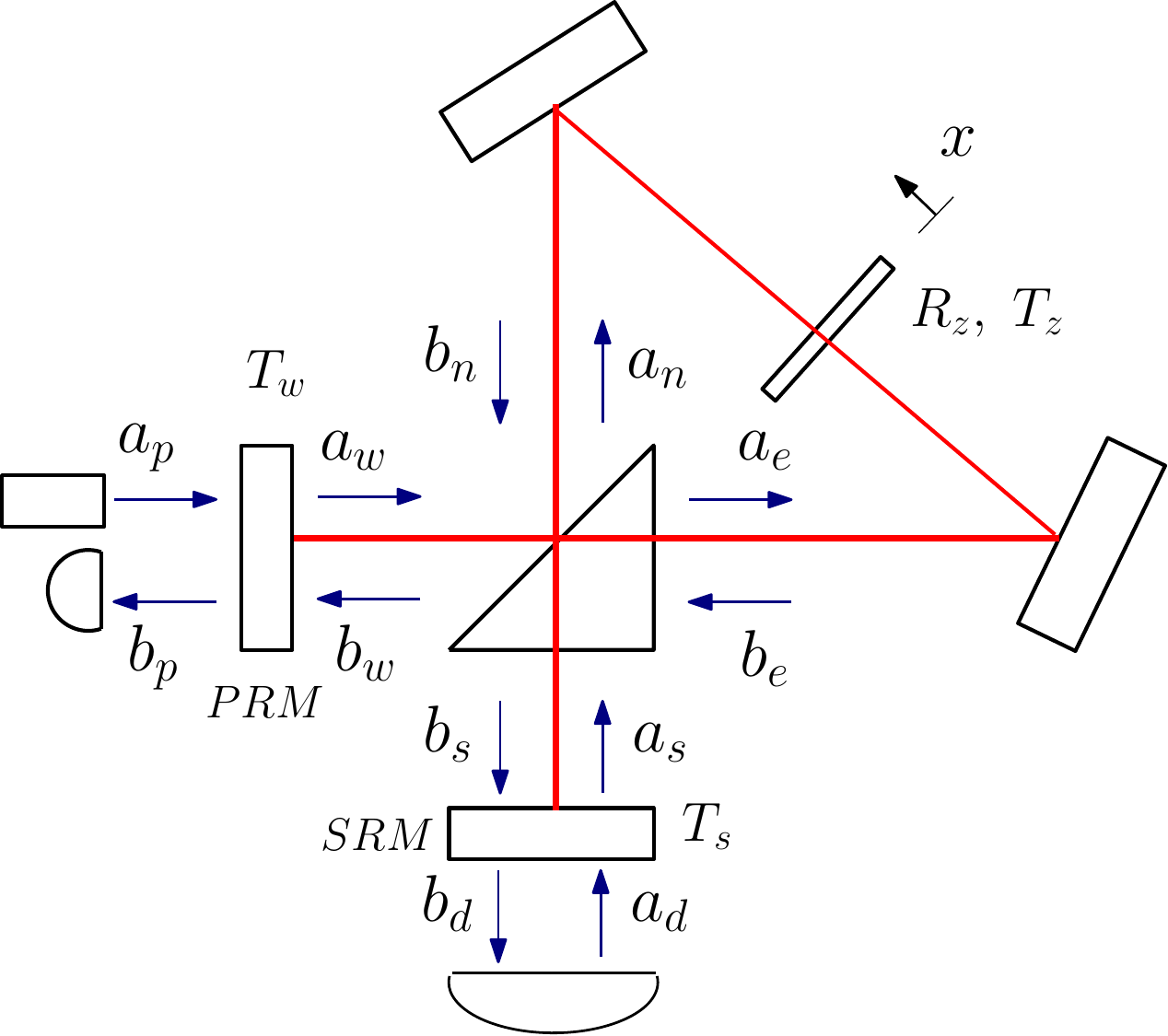}
\caption{Michelson-Sagnac interferometer with Power and Signal recycling mirrors (PRM and SRM). Middle mirror with
amplitude reflectivity $R_z$ may move as a free mass.}\label{ms}
	\end{center}
\end{figure}

DC readout, planned in aLIGO, means introduction of small detuning of arm length. Recall that Michelson 
interferometer with balanced Fabry-Perot (FP) cavities in arms with power and signal recycling mirrors (aLIGO configuration, see  
Fig.~\ref{aligo}) operating in dark port regime possesses  symmetric and antisymmetric  modes, laser pumps symmetric mode and no mean intensity appears in signal (dark) port through signal recycling mirror (SRM). In case of slightly detuned arms small mean intensity appears in signal port. This intensity is used as very stable local oscillator.

The natural question is what {\em arbitrary} (not small) detuning in arms may give for stability. This question became interesting especially after paper of Tarabrin with colleagues  \cite{Tarabrin13} demonstrated possibility of stable optical spring in  Michelson-Sagnac 
interferometer with movable membrane \cite{M-S1, M-S2, M-S3}. Analyzed interferometer with signal recycling mirror (SRM) but without power recycling (PRM) was pumped through power port \cite{Tarabrin13} --- similar configuration is shown on Fig.~\ref{ms} (but with PRM).  However, stability of optical spring was shown for relatively large detuning --- it means relatively 
large power in signal port, which is not convenient in experiment.  Operation far from dark port regime additionally creates the problem of laser noises leaking into signal port --- it makes difficult application of these results to GW detector.

The aim of this paper is to analyze and to demonstrate stable optical rigidity in aLIGO (or Michelson-Sagnac interferometer 
with PRM and SRM) a) pumped throwgh PRM, b) with arm detuning as small as possible (hence, small output power throwgh SRM).
This result may be applied not only to  large-scale gravitational-wave detectors~\cite{Khalili2010} but also to other
optomechanical systems like micromembranes
inside optical cavities ~\cite{Thompson2008} (see Fig.~\ref{ms}), microtoroids~\cite{Verhagen2012},
optomechanical crystals~\cite{Eichenfield2009}, pulse-pumped optomechanical cavities~\cite{Vanner2011}. In spite of the fact that optical rigidity, introduced into micromechanical oscillators, is relatively small as compared with intrinsic one \cite{M-S1}, it may be used for control and manipulation of its dynamics.



\section{Description of model}

We consider a gravitational-wave antenna aLIGO shown on Fig.~\ref{aligo},  amplitude transmittances of SRM  and PRM are
$T_s$ and $T_w$ correspondingly. Antenna consists of a Michelson interferometer with additional mirrors forming Fabry-Perot
(FP) cavities with mean distance $L$ between mirrors in arms which is much larger than distances $\ell$ between beam splitter
and SRM or PRM. Input mirrors have amplitude transmittance $T$ and masses $m$, end mirrors have the same masses $m$ and are
completely reflective. Input and end mirrors in arms may move as free masses. We assume that all mirrors are lossless. The
interferometer is pumped by laser through PRM.


 Recall dynamics of {\em pure balanced} interferometer (i.e. identical FP cavities in arms tuned in resonance with pumped laser) can be split into two modes: namely symmetric and antisymmetric ones. Each mode is characterized by optical detuning $\delta_w$ ($\delta_s$) and decay rate
$\gamma_w$ ($\gamma_s$) dependent on  displacement and transparency of PRM for symmetric mode (SRM for antisymmetric one
correspondingly). Here and below we denote detuning as difference between laser frequency $\omega_0$ and eigen frequencies
$\omega_{w,s}$ of symmetric and antisymmetric modes:
\begin{align}
\label{dwds}
 \delta_w &= \omega_0-\omega_w,\quad  \delta_s = \omega_0-\omega_s\,.
\end{align}
(In aLIGO PRM detuning $\delta_w$ is assumed to be zero, however, below we reserve possibility to vary it.)
The optical fields in the modes represent difference ($e_-$) and sum ($e_+$)
of the fields in arms respectively and carry information about difference ($z_-$) and sum ($z_+$) between lengths of arm
cavities:
\begin{align}
    e_\pm &=\frac{e_e\pm e_n}{\sqrt 2},\\
    z_\pm &= \frac{z_e\pm z_n}{2} ,\quad z_{e,n} \equiv x_{e,n}- y_{e,n},
\end{align}
(see notations on Fig,~\ref{aligo}). In turn, light pressure force may be devided into two part:  fluctuational one
responsible for fluctuational back action and regular part creating optical spring \cite{12a1KhDaLRR}. Below we analyze the
simplified case when sum mechanical displacement is fixed (for example, by feedback):
\begin{align}
    z_+ &=0
\end{align}

When FP cavities in arms are detuned by $\pm \delta$ symmetric and antisymmetric modes became coupled with each other. In this case detunings $\delta_w,\ \delta_s$ \eqref{dwds} and decay rates $\gamma_w,\gamma_s$ refer to partial modes. 
As a result, the system is described by linear set of equations for Fourier components of fields $e_\pm(\Omega),\ e_\pm
^\dag(-\Omega)$ and displacement $z_-(\Omega)$:
\begin{widetext}
    \begin{subequations}
    \label{setEV1}
    \begin{align}
   	(\gamma_w -i\delta_w-i\Omega)\, e_+(\Omega) - i\delta\, e_-(\Omega) - \frac{ik}{\tau}\,E_-z_-(\Omega)
   		&= \frac{\sqrt{\gamma_w}\,g_p(\Omega)}{\sqrt \tau} 	,\\ 
  	-i\delta \, e_+ (\Omega) + (\gamma_s - i\delta_s -i\Omega)\,  e_-(\Omega)  - \frac{ik}{\tau}\, E_+z_-(\Omega) 
  		& = \frac{\sqrt{\gamma_s}\,g_d(\Omega)}{\sqrt \tau},\\ 
  	(\gamma_w +i\delta_w-i\Omega)\, e_+^\dag(-\Omega) + i\delta\, e_-^\dag(-\Omega) + \frac{ik}{\tau}\,E_-^*z_-(\Omega)
   		&= \frac{\sqrt{\gamma_w}\,g_p^\dag(-\Omega)}{\sqrt \tau} 	,\\ 
  	i\delta  e\,_+^\dag(-\Omega) + (\gamma_s +i\delta_s-i\Omega)\,  e_-^\dag(-\Omega)  + \frac{ik}{\tau}\, E_+^*z_-(\Omega) 
  		& = \frac{\sqrt{\gamma_s}\,g_d^\dag(-\Omega)}{\sqrt \tau}\,,\\
  	\hslash k\big\{E^*_+ e_-(\Omega)+ E^*_-e_+(\Omega) + E_+ e_-^\dag(-\Omega)+ E_-e_+^\dag(-\Omega)\big\} +\mu \Omega^2z_-(\Omega) &=0\, ,\\
  	\label{xi}
  	k\equiv \frac{\omega_0}{c},\quad \tau \equiv \frac{L}{c},\quad \mu \equiv \frac{m}{2},\quad E_- \equiv \xi E_+,
  	\quad \xi\equiv \frac{i\delta}{\gamma_s -i\delta_s},\quad  I_+&= \hslash \omega_0 |E_+|^2. 
    \end{align}
    \end{subequations}
\end{widetext}
Here $\hslash$ is  Plank constant, $k$ is wave vector corresponding to laser wave frequency $\omega_0$, $c$ is speed of
light.  $E_\pm$ are mean complex amplitudes of symmetric and antisymmetric modes (excited by pump laser), $I_+$ is power
circulating in symmetric mode.  The right parts ($g_{p,d}(\Omega),\ g_{p,d}^\dag(-\Omega)$) in set describes zero fluctuational
fields incoming into interferometer through PRM and SRM. Details of notations and derivation are presented in
Appendix~\ref{app1}.   

In spite of the fact that set \eqref{setEV1} is not convenient for analysis of sensitivity (because we have to recalculate
fields  $e_\pm$ into output field in signal port), however, it is convenient for optical rigidity analysis.

Following oscillations theory advises we rewrite \eqref{setEV1} introducing normal coordinates $b_\pm(\Omega), \ b_\pm^\dag(-\Omega)$ for e.m. fields and new (complex) eigen values  $\lambda_\pm$:
\begin{subequations}
\label{bpbm}
\begin{align}
 \big(-i\Omega -\lambda_+) b_+ (\Omega) -i z_-\left[\xi - \varkappa\right] &=0\\
  \big(-i\Omega -\lambda_-) b_-(\Omega) - i z_-\big[1+\varkappa\xi\big]& = 0,\\
   \big(-i\Omega -\lambda_+^*) b_+^\dag(-\Omega)  
    +i z_-\left[\xi^* - \varkappa^*\right] &=0\\
  \big(-i\Omega -\lambda_-^*) b_-^\dag(-\Omega)  - i z_-\big[1+\varkappa^*\xi^*\big]& = 0,\\
   \label{f2}
  	\frac{b_+(\Omega)\big[ \xi^*-\varkappa\big] + b_-(\Omega)\big[1 + \xi^*  \varkappa\big]}{d}+ 
  	&\\
  	+ \frac{b_+^\dag(-\Omega)\big[ \xi+\varkappa^*\big] +
		+ b_-^\dag(-\Omega)\big[1 + \xi \varkappa^*\big]}{d^*} + &
		\frac{\Omega^2}{J_+} z_- =0\, ,\nonumber\\
    b_+(\Omega) =  \sqrt \frac{\hslash L^2}{\omega I_+}\left[ e_+(\Omega) -\varkappa e_-(\Omega)\right],&\quad \\
    b_-(\Omega)=  \sqrt \frac{\hslash L^2}{\omega I_+} \left[\varkappa  e_+(\Omega) +  e_-(\Omega)\right].&
\end{align}
\end{subequations}
 
Here we introduce the following notations:
\begin{align}
   \lambda_\pm &= -\left(\Gamma_+ \pm \Gamma_-\sqrt{1 + \Delta^2}\right),	 \quad
  J_+\equiv \frac{k I_+}{L\mu},\\
  \Gamma_\pm &\equiv \frac{ \gamma_w- i\delta_w \pm (\gamma_s -i\delta_s)}{2},
    \quad d\equiv 1+\varkappa^2,\\
    \label{kappa}
  \varkappa & \equiv \frac{i\delta}{\Gamma_w +\lambda_-} 
	  =\frac{\Delta}{1 + \sqrt{1+\Delta^2} }, \quad
	  \Delta \equiv \frac{i\delta}{\Gamma_-}.
\end{align}
In set \eqref{bpbm} we omit fluctuational fields in right parts as we are interesting in {\em dynamic} behavior of system, i.e. in eigen values of determinant.  

After substitution $(-i\Omega\to \lambda$) characteristic equation of set \eqref{bpbm} may be written in form:
  \begin{align}
   \label{ChEqDiff0}
    \lambda^2 
	+\frac{\mathcal I_1\big[1+\alpha_1(\lambda+\tilde \gamma_s)\big]}{(\lambda +\tilde\gamma_s)^2 + \tilde \delta_s^2}
	+\frac{\mathcal I_2\big[1+\alpha_2(\lambda+\tilde \gamma_w)\big]}{(\lambda +  \tilde\gamma_w)^2 +\tilde\delta_w^2}
	  =0\,,
  \end{align}
  where we introduce the following notations:
  \begin{subequations}
  \label{I12}
  \begin{align}
    \tilde \gamma_{w,s} &\equiv -\Re \lambda_\pm,\quad
	  \tilde \delta_{w,s} \equiv \Im \lambda_\pm,\\
	\mathcal I_1 &\equiv  \frac{ 2J_+ \tilde\delta_s\,\Re\phi}{|d|^2}, \quad
		\alpha_1 \equiv \frac{\Im \, \phi}{\tilde\delta_s\,\Re\phi},\\
	\label{I2}
	\mathcal I_2 &\equiv  \frac{ 2J_+ \tilde\delta_w\,\Re\psi}{|d|^2},\quad
		\alpha_2 \equiv \frac{\Im \, \psi}{\tilde\delta_w\,\Re\psi},\\
	\phi &\equiv (1 + \xi^*\varkappa) (1+\varkappa \xi)d^*,\\
	\label{psi}
	\psi &\equiv (\xi^*- \varkappa) (\xi- \varkappa ) d^*\,.
    \end{align}
  \end{subequations}  
  The form of equation \eqref{ChEqDiff0} is the same as for double pumped optical spring \cite{08ChenPRD,2011_RaHiVy_PRD}:
two fractions ($\sim \mathcal I_1$ and $\sim \mathcal I_2$) are similar to two optical springs created in two optical modes
pumped separately. This analogy has physical sense --- for imbalanced interferometer one pump excites two normal modes. This
analogy became more obvious when relaxation rates of symmetric and antisymmetric modes are equal  ($\gamma_w=\gamma_s$). In
this case the values $\varkappa$ and $\xi$ are pure real  and $\alpha_1 = \alpha_2 = 0$. Then characteristic equation takes the following form:
\begin{align}
\lambda^2 
	+\frac{\mathcal I_1}{(\lambda +\tilde\gamma_s)^2 + \tilde \delta_s^2}
	+\frac{\mathcal I_2}{(\lambda +  \tilde\gamma_w)^2 +\tilde\delta_w^2}
	  =0
\end{align}



Note that practically the same set as \eqref{setEV1} is valid for Michelson-Sagnac interferometer shown on Fig.~\ref{ms}
--- see details in Appendix~\ref{app2}. In particular, the equation \eqref{ChEqDiff0} is valid after following
substitutions:
  \begin{align}
  \label{MSIa}
   \delta^2 &\to R_z^2 \delta ^2,\quad J_+ \to R_z^2 J_+, \quad \mu \to m,
  \end{align}
where $R_z$ is amplitude reflectivity of middle mirror, $m$ is its mass.

  \section{Analysis}
  
 Eq.\eqref{ChEqDiff0} may be written in form  convenient for further approximation
 \begin{align}
    \label{ChEqDiff0b}
    D_1^{(0)}& D_2^{(0)} +D^{(1)}=0,\\
    D_1^{(0)}& = \big[ \lambda^2\big((\lambda +\tilde\gamma_s)^2 + \tilde \delta_s^2\big) 
	  +\mathcal I_1\big(1+\alpha_1(\lambda+\tilde \gamma_s)\big],\\
    D_2^{(0)} & = \big[(\lambda +  \tilde\gamma_w)^2 +  \tilde\delta_w^2\big],\\
    D^{(1)} &= \big[(\lambda +  \tilde\gamma_s)^2 +  \tilde\delta_s^2\big]
	  \mathcal I_2\big(1+\alpha_2(\lambda+\tilde \gamma_w)\big).
    \end{align}

Underline that Eq. \eqref{ChEqDiff0b} is still exact characteristic equation. Mathematically its left part is a polynomial of
$6$-th degree relatively variable $\lambda$. Its solution provides set of eigenvalues $\lambda_k$, its imaginary parts
describe eigen frequencies whereas real parts -- relaxation rates (positive one corresponds to instability). It is not
difficult task for numerical solution of \eqref{ChEqDiff0b} using contemporary mathematical packets. However, 
analysis based on numeric calculations is not simple because there is set of $6$  parameters ($\gamma_{w,s},\ \delta_{w,s},\ \delta,\ I_+$) which may be varied.  

In theoretical  analysis below  we make following assumptions:
\begin{itemize}
 \item Interferometer is pumped through PRM.
 \item Arm detuning is small: $\delta\ll \delta_{w,s}$.
 \item Initial relaxation rates are small: $\gamma_{w,s} \ll \delta_{w,s}$.
\end{itemize}
Then Eq.~\eqref{ChEqDiff0b} may be solved by iteration method considering  term $D_1^{(0)} D_2^{(0)}$ as main term (in zero
approximation roots are  $\lambda_k^{(0)}$), whereas account of term $D^{(1)}$ of first order of smallness gives next
iteration  $\lambda_k^{(1)}$.  We can do that because coefficients $\xi,\ \varkappa \sim \delta$  (\ref{xi}, \ref{kappa}), hence, $\psi \sim \delta^2$ \eqref{psi} and the "additional" pump $\mathcal I_2\sim \delta^2$ \eqref{I2}. It means that  $\mathcal I_2$ is much smaller than the "main" pump $\mathcal I_1$ and we may apply iteration method.

\paragraph*{Zero order iteration.} The solution of equation $D_1^{(0)}=0$ is following:
\begin{align}
  \label{lambda12}
   & \lambda_{1,2}^{(0)} = \gamma_1 \pm i \delta_1,\quad \lambda_{3,4}^{(0)} = \gamma_3 \pm i \delta_3\\
    \gamma_1& \equiv \frac{\tilde\gamma_s (1 - p) -(1-p^2)\beta_1}{ 2 p},\quad
    \delta_1 \equiv \sqrt{ \frac{\tilde\gamma_s^2 + \tilde \delta_s^2}{2}\big(1 - p\big)}\,,\nonumber\\
    \label{lambda34}
 \gamma_3 &\equiv -\frac{\tilde\gamma_s (1 + p) 
		-(1-p^2)\beta_1}{2 p},\quad
    \delta_3 \equiv \sqrt{ \frac{\tilde\gamma_s^2 + \tilde \delta_s^2}{2}\big(1 + p\big)}\,,\nonumber\\
    p&\equiv \sqrt{1- \frac{ 4\mathcal I_1\big(1 + \alpha_1 \tilde \gamma_s\big)}{
		\big[\tilde\gamma_s^2 + \tilde \delta_s^2\big]^2} },\quad 
	\beta_1\equiv \frac{\alpha_1\big(\tilde \gamma_s^2 +\tilde \delta_s^2\big)}{
		4\big(1 + \alpha_1 \tilde \gamma_s\big)}
\end{align}
Note that in case of zero arm detuning ($\delta=0$) these roots was found earlier \cite{01a1KhPLA,05a1LaVyPLA,06a1KhLaVyPRD}
(for example, the case of $p=0$ corresponds to double resonance regime) and formulas above may be considered as
generalization for small $\delta$. 

Solution of equation $D_2^{(0)}=0$ gives obvious roots:
\begin{align}
    \lambda_{5,6}^{(0)} &=- \tilde \gamma_w \pm i \tilde \delta_w
\end{align}

So in zero order approximation we have roots $\lambda_k^{(0)}$, among them the roots $\lambda_{1,2}^{(0)}$ correspond to instability ($\gamma_1>0$). Now zero order part of determinant may be written as

\begin{align}
  \label{ChEqDiff1}
    D_1^{(0)} D_2^{(0)} &= \big[(\lambda-\gamma_1)^2+\delta_1^2\big]\big[(\lambda-\gamma_3)^2+\delta_3^2\big]\times\\
	  &\qquad \times\big[(\lambda +\tilde \gamma_w)^2 +\tilde \delta_w^2\big].\nonumber
\end{align}

\paragraph*{First order of iteration.} Our aim is to choose such parameters which make stable next iteration root $\lambda_{1,2}^{(1)}$, i.e. 
\begin{align}
    \Re \left[ \lambda_{1,2}^{(1)}\right] < 0
\end{align}
We divide \eqref{ChEqDiff0b} by $\big[(\lambda-\gamma_3)^2 +\delta_3^2\big]$ (taking into
account \eqref{ChEqDiff1}) and put $\lambda=\lambda_{1,2}^{(0)}$ in $D^{(1)}$. So we get next iteration of characteristic
equation:
\begin{align}
 \label{ChEqNe}
    \big((\lambda -\gamma_1)^2& +\delta_1^2\big) \big((\lambda +\tilde \gamma_w)^2 + \tilde \delta_w^2\big) -b=0,\\
	b &\equiv - \left.\frac{D^{(1)}}{(\lambda -\gamma_3)^2 +\delta_3^2 }\right|_{\lambda= \lambda_{1,2}^{(0)}}\,.
\end{align}
We may keep in mind $b$ as a constant of first order of smallness.

Below we put  $ \tilde\delta_w \simeq -\delta_1$, it is this choose of $\tilde\delta_w$ that provides stability with
minimal arm detuning $\delta$. This choice has physical sense corresponding to known scheme of laser cooling (see, for example \cite{coolingV, cooling}). Indeed, let
FP cavity, which one mirror is a mechanical oscillator with frequency $\omega_m$, is pumped by laser with
frequency {\em less} than cavity frequency by $\omega_m$ detuned from resonance. In this case positive damping will be
created for movement of mechanical oscillator (optical rigidity is negligibly small).

One may write down solution
of \eqref{ChEqNe} in analytical form
\begin{align}
 \label{lambdaNe}
 \lambda &=\frac{\gamma_1- \tilde \gamma_w}{2}\pm \\
    &\quad \pm i\sqrt{\delta_1^2 - \left[\frac{\gamma_1+\tilde \gamma_w}{2}\right]^2 
	  \pm \sqrt{b -\delta_1^2\big[\gamma_1+ \tilde \gamma_w \big]^2}} .\nonumber
\end{align}
Analysis shows that $\Im b\ll \Re b$. Then at condition
\begin{align}
    \label{cond1}
    \Re b= \delta_1^2\big[\gamma_1+ \tilde \gamma_w \big]^2
\end{align}
the second term in \eqref{lambdaNe} is practically imaginary and its real part is small enough. Then the condition stability may be approximately formulated as
\begin{align}
    \label{cond2}
 \tilde\gamma_w > \gamma_1,\quad \text{or }\ \tilde\gamma_w > \tilde\gamma_s\frac{1-p}{2p}
\end{align}

This conditions  give an estimation for minimal value of arm detuning:
\begin{align}
\label{detun}
\delta^2 > &[\tilde{\gamma}_s \frac{1 - p}{2p}+ \tilde{\gamma}_w]^2 \times\\
&\times \left( \frac{\sqrt{2}+\sqrt{1-p}}{2\sqrt{2}+\sqrt{1-p}}\right)^2 \frac{4\sqrt{2}p}{(1-p)^{1/2}(1+p)^2} \nonumber
\end{align}
The formula \eqref{detun} is confirmed by numerical calculations presented in the following section. 

 Important that  in order to fulfill condition \eqref{detun} one has to provide relatively small arm detuning $\delta\sim
\tilde\gamma_w$.  Here we made an assumption that $\tilde \gamma_{w,s}$ depend weakly on a value of $\delta$. So we put $\tilde \gamma_{w,s} \simeq \gamma_{w,s}$ correspondingly when doing numerical estimations, because otherwise \eqref{detun} turns into non-trivial equation for $\delta$ (we did this approximation only estimating value of $\delta$, other numerical calculations stay exact).

\begin{table}[h] 
\caption{\label{aLIGO}Parameters for aLIGO}
\begin{center}
\begin{tabular}{|c|c|}
\hline
Detuning of symmetric mode ($\delta_w$) &  -23.0 Hz\\
\hline
Detuning of antisymmetric mode ($\delta_s$) & 42.4 Hz \\
\hline
Decay rate of symmetric mode ($\gamma_w$)& 1.5 Hz (3.0 Hz) \\
\hline
Decay rate of antisymmetric mode ($\gamma_s$) & 0.3 Hz (3.0 Hz) \\
\hline
Test mass (m) & 40 kg \\
\hline
Arm length (L) & 4 km \\
\hline
Circulating power ($I_{circ}$)& 24 kW \\
\hline
Arm detuning ($\delta$) &  1.51 Hz (4.6 Hz)\\
\hline
\end{tabular}
\end{center}
\end{table}

\section{Numerical estimations}
\begin{figure}[t]
\includegraphics[width=0.35\textwidth]{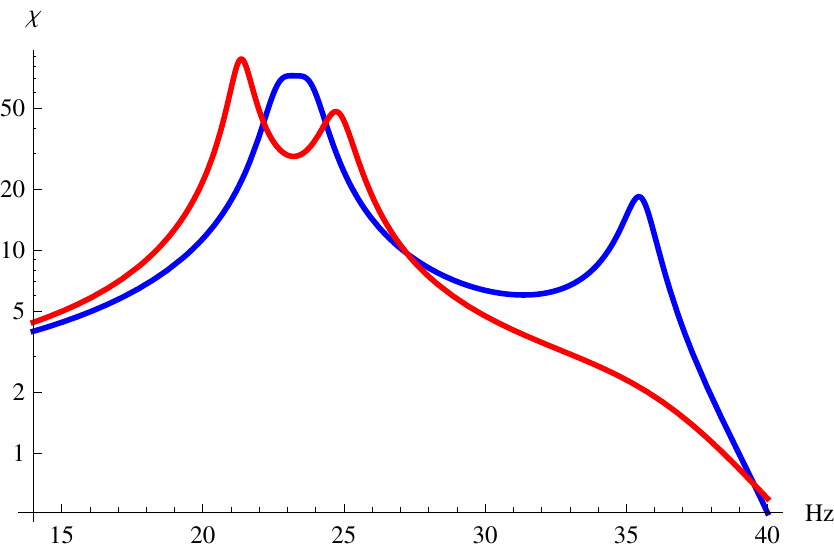}
\caption{Susceptebility $\chi$ of aLIGO interferometer with excluded symmetric mechanical mode. Red curve -- $\tilde{\gamma}_w = \tilde{\gamma}_s$. Blue curve -- $\tilde{\gamma}_w \neq \tilde{\gamma}_s$ \label{susc1}}
\includegraphics[width=0.35\textwidth]{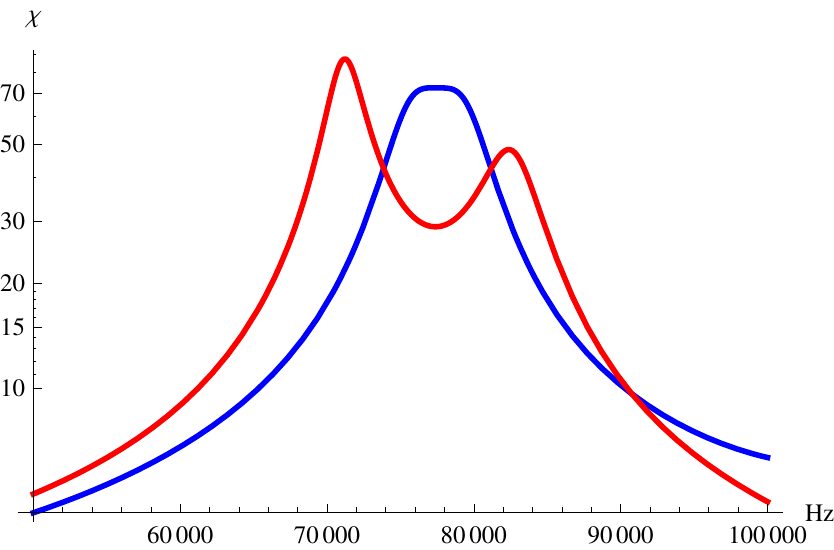}
\caption{Susceptebility $\chi$ of Michelson-Sagnac interferometer. Red curve -- $\tilde{\gamma}_w = \tilde{\gamma}_s$. Blue curve -- $\tilde{\gamma}_w \neq \tilde{\gamma}_s$ \label{susc2}}
\includegraphics[width=0.35\textwidth]{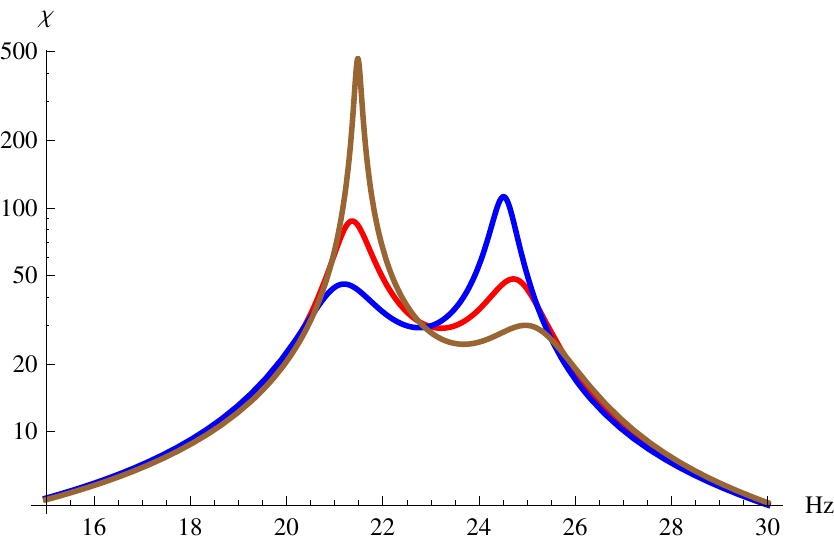}
\caption{Susceptebility $\chi$ of aLIGO with $\tilde \delta_w = - \delta_1 + \Delta$. Red curve -- $\Delta = 0$. Blue curve -- $\Delta = 0.5 ~Hz$. Brown curve -- $\Delta = - 0.5~ Hz$ \label{susc3}}
\end{figure}

Numerical estimations can serve as an examination of our theory. We can solve \eqref{ChEqDiff0b}  numerically substituting realistic parameters. We chose the parameters for aLIGO interferometer presented in Table~\ref{aLIGO} \cite{aLIGOref}. We consider two cases --- when $\tilde \gamma_w \neq \tilde \gamma_s$ and when $\tilde \gamma_w = \tilde \gamma_s$. In a table values in brackets mean second case. Our analysis also gives an estimation for output power $I_{out} = 0.03 W$ (as we know in aLIGO reference design output power should be about $0.1 W$). It is a good result because we don't want to obtain big laser power on a photodetector. Importantly that here we chose operating frequency about 30 Hz. This value differs from aLIGO one --- 100 Hz. We made it because in our case power-recycling mirror is detuned from resonance. From this fact follows that the circulating power ($\sim$ 24 kW) is less than in aLIGO ($\sim$ 800 kW). Susceptibility curves for this parameters are represented on a Fig.
\ref{susc1}. Numerical solution of \eqref{ChEqDiff0b} gives us a set of eigenvalues with negative real parts --- it means stability. Important that numerical egenvalues are in good agreement with analytical estimates. In addition we checked our analysis numerically by Routh -- Hurwitz stability criterion. It showed stability for parameters predicted by our 
theory.

We also did the same analysis for Michelson-Sagnac interferometer. For such system we chose  realistic parameters presented in Table~\ref{MSI} \cite{Tarabrin13, M-S1}. However, we consider membrane as a free mass not taking into account its intrinsic rigidity. Numerical solution gives us a set of eigenvalues with negative real parts again. Plots of susceptibilities are represented on a Fig.\ref{susc2}.

\begin{table}[h] 
\caption{\label{MSI}Parameters for Michelson-Sagnac interferometer}
\begin{center}
\begin{tabular}{|c|c|}
\hline
Detuning of symmetric mode ($\delta_w$) &  -77.2 kHz\\
\hline
Detuning of antisymmetric mode ($\delta_s$) &  141.0 kHz \\
\hline
Decay rate of symmetric mode ($\gamma_w$)& 5 kHz (10 kHz) \\
\hline
Decay rate of antisymmetric mode ($\gamma_s$) & 1 kHz (10 kHz)\\
\hline
Test mass (m) & $10^{-10}$ kg \\
\hline
Arm length (L) &8.7 cm \\
\hline
Circulating power ($I_{circ}$)& 318 mW \\
\hline
Arm detuning ($\delta$) &  5 kHz (15 kHz)\\
\hline
Membrane reflectivity ($R_z^2$) & 0.17\\
\hline
\end{tabular}
\end{center}
\end{table}

Our analysis shows that we can control the shape of the susceptibility curve (increase  one peak and decrease another one) just detuning $\tilde \delta_w$  by small value $\Delta$ from optimal one: $\tilde \delta_w=-\delta_1 + \Delta$. On Fig.~\ref{susc3} we plot such curves for parameters represented in Table. \ref{aLIGO}.


\section{Conclusion}

We have shown that arm detuning $\delta$ in aLIGO interferometer provides possibility to make {\em stable} optical spring for
antisymmetric mechanical mode. Important that the stable optical spring  may be created with  {\em small} arm detuning
comparable with optical bandwidths: $\delta\simeq \gamma_w,\ \gamma_s$. However, this regime requires relatively large PR and SR detunings which  restrict power circulating in arms of
interferometer.

This results may be easy applied to table top Michelson-Sagnac interferometer with membrane inside to create stable optical
spring.


We restrict ourselves by analysis of {\em only antisymmetric} mechanical mode in detuned aLIGO interferometer.
In further research we plan answer on question: is it possible to make {\em both symmetric and antisymmetric} mechanical modes to
be stable?

\acknowledgments
We are grateful to R. Adhikari, Y.~Chen,  H.~Miao, F.~Khalili and especially to S.~Tarabrin for fruitful discussions.
Authors are supported  by the Russian Foundation for Basic Research Grants No. 08-02-00580-a, 13-02-92441 and NSF grant PHY-1305863.


\appendix
\section{Notations and derivation of initial equations}
\label{app1}


Here we explain  notations and derive set of equations \eqref{setEV1}, describing aLIGO scheme represented on a Fig.\ref{aligo}. 

Electrical field $E$ of optical wave is presented in a standard way: 

\begin{align}
\label{fields1}
E &=  \sqrt{\frac{2 \pi \hbar \omega_0}{S c}}e^{- i \omega_0 t}\left( A  
	  + a_\text{fl} \right)  + \text{h.c.}\\
    a_\text{fl} &= \int_0^{\infty} \sqrt\frac{\omega}{\omega_0} \,a(\omega) e^{-i (\omega-\omega_0) t} \frac{d \omega}{2 \pi}
    \nonumber
\end{align}

where $A$ is mean amplitude, $\omega_0$ is mean frequency, (mean power $P$ of traveling wave is $ P=\hslash \omega_0 |A|^2$),  $a(\omega)$ -- operators describing quantum fluctuations, their commutators are
\begin{align}
\label{commutators}
 \big[a(\omega),\, a^\dag(\omega')\big] & = 2\pi\, \delta(\omega-\omega').
\end{align}
Usually fluctuation part is written in form: 

\begin{align}
\label{fields2}
a_\text{fl} &\simeq  
	  \int_{-\infty}^{\infty} {a}(\Omega) e^{-i \Omega t} \frac{d \Omega}{2 \pi}\, ,
\end{align}
where $\Omega= \omega-\omega_0$  (see details in \cite{02a1KiLeMaThVyPRD}).
We assume that input wave is in coherent state. In this case we have for averages:
\begin{align}
 \big\langle a(\Omega)\, a^\dag(\Omega')\big\rangle & = 2\pi\, \delta(\Omega-\Omega'),\
 \big\langle  a^\dag(\Omega')\,a(\Omega)\big\rangle  = 0
\end{align}

In our notations we use big letters for mean (classical) part of field and small letters for small additions including quantum fluctuating component. 

\subsection{The beamsplitter}

For incident and reflected fields on beam splitter we assume following formulas  
\begin{subequations}\label{eil_bs_raw}
  \begin{gather}
    b_w = - \frac{b_e+b_n}{\sqrt2} \,, \qquad
    a_b = - \frac{b_e-b_n}{\sqrt2} \,, \\
    a_e = - \frac{a_w+a_s}{\sqrt2} \,, \qquad
    a_n = - \frac{a_w-a_s}{\sqrt2} \,.
  \end{gather}
\end{subequations}

\subsection{Mean fields}
For reflected fields of east and north cavities we can write:
\begin{align}
B_e&={\cal R}_e A_e,\quad B_n={\cal R}_n A_n,
\end{align}

The both east and north arms are assumed to be slightly detuned by $\delta$ from resonance to  opposite sides. We introduce following notations and calculate generalized reflectivities ${\cal R}_e, \ {\cal R}_n$ in long way approximation:
\begin{align}
\Theta_e&= e^{-i\delta\tau},\quad \Theta_n = e^{i\delta\tau},\
	\delta=\omega_0-\omega_{res},\\
\gamma_T& =\frac{T^2}{4\tau} ,\ \tau=\frac {L}{c},\nonumber\\
{\cal R}_e & \equiv \frac{\gamma_T+i\delta}{\gamma_T -i\delta}= {\cal R}_n^*,\quad 
	{\cal R}_n\equiv \frac{\gamma_T-i\delta}{\gamma_T +i\delta}\,.
\end{align}

Using (\ref{eil_bs_raw}) we get
\begin{subequations}
\label{eil_PM_0}
  \begin{align}
  A_e &=-(A_w+A_s)/\sqrt 2,\quad   B_e=\mathcal R_e A_e\\
  A_n &= -(A_w-A_s)/\sqrt 2,\quad   B_n=\mathcal R_n A_n\\
  B_w &= -(B_e+B_n)/\sqrt 2=A_w\mathcal R_+ +A_s\mathcal R_- \,,\\
   B_s &= -(B_e-B_n)/\sqrt 2= A_w\mathcal R_- + A_s\mathcal R_+ ,
     \end{align}
\end{subequations}
where we introduced $\mathcal R_\pm \equiv \frac{\mathcal R_e \pm \mathcal R_n}{2}\,$,
\begin{subequations}
   \begin{align}
    \label{Rpm}
    \mathcal R_+ = \frac{\gamma_+\gamma_- -\delta^2}{\gamma_+^2+\delta^2},\quad 
    	\mathcal R_- = \frac{i\delta (\gamma_+ + \gamma_-)}{\gamma_+^2+\delta^2}.
  \end{align}
\end{subequations}

Now we may consider the SR (south) and PR (west) cavities which are described by equations (keep in mind that there is no pumping into the south arm, but keeping $A_d$ yet):
\begin{subequations}\label{eil_WS_0}
  \begin{align}
    &B_p = -R_wA_p + iT_w\Theta_wB_w \,, \\
    \label{eil_WS_0b}
    &A_w = -R_w\Theta_w^2B_w + iT_w\Theta_w A_p \,, \\
    &B_d = iT_s\Theta_sB_s -R_s A_d\,, \\
    \label{eil_WS_0d}
   & A_s = -R_s\Theta_s^2B_s + iT_s\Theta_sA_d \,,\\
   & \Theta_{w,s} \equiv e^{i\omega_0\tau_{w,s}} \,.
  \end{align}
\end{subequations}

Using (\ref{eil_PM_0}) one may write set of linear equations \eqref{eil_WS_0b}, \eqref{eil_WS_0d} for $A_s$ and $A_w$ which may be solved for non zero $A_d$:
\begin{align}
   A_w&(1+R_w\Theta_w^2\mathcal R_+) + A_s R_w\Theta_w^2\mathcal R_- =iT_w\Theta_w  A_p,\\
   A_w&\,R_s\Theta_s^2\mathcal R_- +A_s(1+R_s\Theta_s^2\mathcal R_+) = iT_s \Theta_s A_d \, .
\end{align}

Solving this set  and simplifying the solution one get:
\begin{align}
\label{Bw}
 A_w &= \frac{iA_p\sqrt{ \gamma_w/\gamma_T}e^{i\alpha_w}
 	\big(\gamma_+\Gamma_s +\delta^2\big)}{
 	\Gamma_s\Gamma_w+\delta^2}-\\ \nonumber
 	&\quad -\,\frac{iA_d\sqrt{ \gamma_s/\gamma_T}e^{i\alpha_s}\,
 		2i\delta \gamma_T\, R_w\Theta_w^2}{\big(\Gamma_s\Gamma_w+\delta^2\big)(1-R_w\Theta_w^2)}\,, \\
\label{Bs}
A_s &= \frac{iA_d\sqrt{\gamma_s/\gamma_T}e^{i\alpha_s}\big(\gamma_+\Gamma_w +\delta^2\big)}{
 	\Gamma_s\Gamma_w+\delta^2}-\\
 	&\quad -\,\frac{iA_p\sqrt{ \gamma_w/\gamma_T}e^{i\alpha_w}\,2i\delta \gamma_T R_s\Theta_s^2}{
 	\big(\Gamma_s\Gamma_w+\delta^2\big)(1-R_s\Theta_s^2)}\,, \nonumber
\end{align}
where we introduced  notations:
\begin{align}
\label{Gamma_sw}
 &\Gamma_{s,w} \equiv
 	\gamma_{s,w} - i\delta_{s,w} \equiv  
 		\frac{\gamma_+ + \gamma_-R_{s,w}\Theta_{s,w}^2}{1-R_{s,w}\Theta_{s,w}^2}	\\
 	&\ \gamma_{s,w} = \frac{\gamma_T(1 - R_{s,w}^2)}{
    	\big|1-R_{s,w}\Theta_{s,w}^2\big|^2} ,\,\\
 & \delta_{s,w} \equiv -\left[\frac{\gamma_++\gamma_-}{2}\right]
     \frac{R_{s,w}\big[\Theta_{s,w}^2-\Theta_{s,w}^{*2}\big]}{
    	\big|1-R_{s,w}\Theta_{s,w}^2\big|^2}\, ,\\
    	& e^{i\alpha_{s,w}} \equiv \frac{\Theta_{s,w}\big|1-R_{s,w}\Theta_{s,w}^2\big|}{1-R_{s,w}\Theta_{s,w}^2}=
    	\sqrt\frac{\Theta_{s,w}^2-R_s}{1-R_{s,w}\Theta_{s,w}^2} \,. \nonumber
\end{align}

Now we can calculate fields before input mirrors in arms using \eqref{Bw}, \eqref{Bs} and \eqref{eil_WS_0}:
\begin{align}
 A_e &= -\frac{i\sqrt{ \gamma_w/\gamma_T}\,A_pe^{i\alpha_w}(\gamma_+ - i\delta)
    	\big(\Gamma_s +i\delta\big)}{
    \sqrt 2\big(\Gamma_s\Gamma_w+\delta^2\big)}-\\ \nonumber
    &\quad - \frac{i\sqrt{ \gamma_s/\gamma_T}\,A_de^{i\alpha_s}
    	(\gamma_+ - i\delta)\big(\Gamma_w +i\delta\big)}{
    \sqrt 2\big(\Gamma_s\Gamma_w+\delta^2\big)}, \\ \nonumber
  A_n &=  - \frac{i\sqrt{ \gamma_w/\gamma_T}\,A_pe^{i\alpha_w}
    	(\gamma_+ + i\delta)\big(\Gamma_s -i\delta\big)}{
    \sqrt 2\big(\Gamma_s\Gamma_w+\delta^2\big)} + \\ 
    &\quad + \frac{i\sqrt{ \gamma_s/\gamma_T}\,A_de^{i\alpha_s}
    	(\gamma_+ + i\delta)\big(\Gamma_w -i\delta\big)}{
    \sqrt 2\big(\Gamma_s\Gamma_w+\delta^2\big)}.
\end{align}
And finally  we calculate mean fields circulating in arms:
\begin{align}
\label{Een}
 E_{e,n} &= \mathcal T_{e,n} A_{e,n},\quad  
 	\mathcal T_{e,n}= \frac{i\sqrt{\gamma_T /\tau}}{\gamma_+ \mp i\delta},\\
 \label{Epm}
 E_\pm & = \big( E_e\pm E_n\big)/\sqrt 2\\
 \label{Ep}
 E_+  = & \frac{\sqrt{\ \gamma_w/\tau}\,A_p e^{i\alpha_w}\Gamma_s }{
    \big(\Gamma_s\Gamma_w+\delta^2\big)} +  
      \frac{\sqrt{ \gamma_s/\tau}\,A_d e^{i\alpha_s} i\delta}{
    	\big(\Gamma_s\Gamma_w+\delta^2\big)}\,,\\ 
 \label{Em}
 E_-  = & \frac{\sqrt{ \gamma_w/\tau}\,A_p e^{i\alpha_w} i\delta }{
    \big(\Gamma_s\Gamma_w+\delta^2\big)} +  
      \frac{\sqrt{ \gamma_s/\tau}\,A_d e^{i\alpha_s} \Gamma_w}{
    	\big(\Gamma_s\Gamma_w+\delta^2\big)}\, . 
\end{align}

\subsection{Small fields}
Below we consider small (and fluctuative) part of a field in frequency domain. The logic of derivation is the same, but in this situation fluctuative part contains information on spectral frequency $\Omega$.

\subsubsection{East and north arms}
We use long wavelength approximation for arm cavity. In particular, we assume that field reflected from arm contains  information on {\em difference} coordinates $z_{e,n}$ of arm.  So we assume that $b_{n,e}$ and $a_{n,e}$ may be expressed by formulas:
\begin{subequations}
 \label{bEN}
\begin{align}
  \label{benaen}
& b_{e,n} = a_{e,n}\,{\mathbb R}_{e,n} -E_{e,n}{\mathbb T}_{e,n}\,  2ikz_{e,n},\\
\label{eenaen}
 & e_{e,n} = a_{e,n}\,{\mathbb T}_{e,n} - E_{e,n}\frac{{\mathbb T}_{e,n}}{iT}\, 2ikz_{e,n},\\
 &\mathbb R_e= \frac{\gamma_T+i(\delta+\Omega)}{ \gamma_T - i(\delta+\Omega)}, \quad
    \mathbb R_ n= \frac{\gamma_T+i(\Omega-\delta)}{ \gamma_T - i(\Omega-\delta)},\,\\
 & \mathbb T_{e}= \frac{i\sqrt{\gamma_T/\tau}}{\gamma_T-i(\delta+\Omega) },\quad
 	\mathbb T_{n}= \frac{i\sqrt{\gamma_T/\tau}}{\gamma_T-i(\Omega - \delta) },\\
 & z_{n,e} =x_{n,e}-y_{n,e}.
\end{align}
\end{subequations}

\subsubsection{Beamsplitter}
Now we may calculate using (\ref{eil_PM_0})
\begin{subequations}
 \label{Defs}
\begin{align}
 a_e &= -\,\frac{a_w+a_s}{\sqrt 2},\quad a_n= -\,\frac{a_w-a_s}{\sqrt 2},\\
 \label{as}
 b_s & = -\,\frac{b_e-b_n}{\sqrt 2}=a_w\mathbb R_- +a_s\mathbb R_+  +  \mathbb Z_s,\\
 \label{aw}
 b_w &= -\,\frac{b_e+b_n}{\sqrt 2}=a_w\mathbb R_+ +a_s\mathbb R_-  +  \mathbb Z_w,
\end{align}
\end{subequations}
where we introduced following notations:
\begin{align}
    \label{Zs}
   & \mathbb Z_s = {\mathbb T}_- W_= + {\mathbb T}_+W_\times \\
   \label{Zw}
    & \mathbb Z_w = {\mathbb T}_+  W_= + {\mathbb T}_- W_\times \,,\\
   & W_= \equiv \big[E_+z_+ + E_-z_-\big]2ik,\,\\
   & W_\times \equiv \big[E_+z_-+ E_-z_+\big]i2 k, \,\\
    &\mathbb T_+\equiv \frac{\mathbb T_e + \mathbb T_n}{2}=
    	\frac{i\sqrt{\gamma_T/\tau}\,(\gamma_T - i\Omega)}{(\gamma_T - i\Omega)^2
 	 +\delta^2},\,\\ 
   &\mathbb T_-\equiv \frac{\mathbb T_e - \mathbb T_n}{2}=
    	\frac{i\sqrt{\gamma_T/\tau}\, i\delta}{(\gamma_T - i\Omega)^2 +\delta^2},\, \\ 
	&z_\pm=\frac{z_e\pm z_n}{2},\quad 
	E_\pm=\frac{E_e\pm E_n}{\sqrt 2}\,.
\end{align}

\subsubsection{Inside fields  in arms }
Fields $e_{e,n}$ inside arms may be calculated using \eqref{eil_PM_0} and \eqref{eenaen}. We may pass trough sum and different fields  $e_{\pm} = \frac{e_e \pm e_n}{\sqrt{2}}$.
Instead of $\big\{ a_d,\, a_p\big\}$ we may inroduce the new basis for fluctuation amplitudes:
\begin{align}
\label{newb}
 g_p & = e^{i\alpha_w} a_p ,
 	\quad
 	g_d  = e^{i\alpha_s} a_d.
\end{align}
The fluctuational amplitudes $\big\{g_p,\, g_d\big\}$ are independent from each other as well as 
$\big\{ a_d,\, a_p\big\}$, i.e. their cross correlators are equal to zero and own correlators are the same as for initial basis (see \eqref{commutators})
\begin{align}
[g_d(\Omega),\, g_d^\dag(\Omega')\big] &=  2\pi \, \delta(\Omega-\Omega'), \\ 
[g_p(\Omega),\, g_p^\dag(\Omega')\big] &= 2\pi \, \delta(\Omega-\Omega') \, .
\end{align}
After simple but bulky calculations we obtain  expressions  for $e_\pm$:
\begin{subequations}
\label{epmNB}
\begin{align}
 \label{e+NB}
 e_+  & = \frac{\big[\Gamma_{s}-i\Omega\big]\,\sqrt{\gamma_w}\,g_p 
 		+ i\delta\,\sqrt{ \gamma_s}\,g_d }{
 		\sqrt{\tau}\big[(\Gamma_s- i\Omega)(\Gamma_w-i\Omega) + \delta^2\big]}
		+  \\ \nonumber
	  & \qquad +\frac{ W_= \big[\Gamma_{s}-i\Omega\big] + W_\times  i\delta }{2\tau
 		\big[(\Gamma_s- i\Omega)(\Gamma_w-i\Omega) + \delta^2\big]},\,\\
 \label{e-NB}
  e_- & = \frac{ i\delta\, \sqrt{ \gamma_w}\,g_p +\big[\Gamma_w -i\Omega\big]\sqrt{ \gamma_s}\,g_d
  		 }{\sqrt{\tau}
 		\big[(\Gamma_s- i\Omega)(\Gamma_w-i\Omega) + \delta^2\big]} + \nonumber \\
       & \qquad + \frac{ W_\times  \big[\Gamma_{w}-i\Omega\big]  + W_= \, i\delta 
	 }{2\tau \big[(\Gamma_s- i\Omega)(\Gamma_w-i\Omega) + \delta^2\big] }\,.
\end{align}
\end{subequations}

We can rewrite formulas \eqref{epmNB} in form:
\begin{align}
\eqref{e+NB} &\times (\Gamma_w - i \Omega) + \eqref{e-NB} \times (- i \delta) \Rightarrow\nonumber\\
\label{char1}
(\Gamma_w - i \Omega)& e_+ - i\delta e_- = \frac{\sqrt{\gamma_w} g_p}{\sqrt{\tau}} + \frac{i k (E_+ z_+ + E_- z_-)}{\tau},\\
\eqref{e-NB} &\times (\Gamma_s - i \Omega) + \eqref{e+NB} \times (- i \delta) \Rightarrow,\nonumber\\
\label{char2}
-i \delta e_+& + (\Gamma_s - i\Omega)e_- = \frac{\sqrt{\gamma_s} g_d}{\sqrt{\tau}} + \frac{ik (E_+ z_- + E_- z_+)}{\tau}.
\end{align}
Equations \eqref{char1} and \eqref{char2} (and their complex conjugation) form first four equations of set \eqref{setEV1} if we exclude symmetric mode (putting $z_+ = 0$).

\subsubsection{Ponderomotive forces and equations of motion}
We can express forces acting on end mirror in each arm in next way:
\begin{align}
 F_{e,n} & = 2\hslash k \big(E_{e,n}^*e_{e,n}+E_{e,n}e_{e,n}^\dag\big),\\
 F_+ & \equiv \frac{F_e + F_n}{2},\quad
 F_-   \equiv \frac{F_e- F_n}{2}
\end{align}
After that we can write equations of motion for symmetric and antisymmetric modes in frequency domain:
\begin{align}
\label{mot1}
\hbar k \big[ E_+^*e_-(\Omega) + E_-^*e_+(\Omega) +\big\{\text{h.c.}\big\}_- \big] + \mu \Omega^2 z_-(\Omega) = 0, \\
\label{mot2}
\hbar k \big[E_+^*e_+(\Omega)+E_-^*e_-(\Omega) +\big\{\text{h.c.}\big\}_- \big] + \mu \Omega^2 z_+(\Omega) = 0\, .
\end{align}
Equation \eqref{mot1} forms last equation of set \eqref{setEV1}.

\section{ Comparison of Michelson and Michelson-Sagnac interferometers}\label{app2}

Here we prove the formulas \eqref{MSIa}. We consider simplified Michelson interferometer on Fig.~\ref{Mich}, show that it is similar to aLIGO interferometer and it is described by set similar to \eqref{setEV1}. Then we consider Michelson-Sagnac interferometers and compare it with Michelson interferometer.
\subsection{Michelson interferometer}
\begin{figure}[h]
	\begin{center}
		 \includegraphics[width = .85\linewidth]{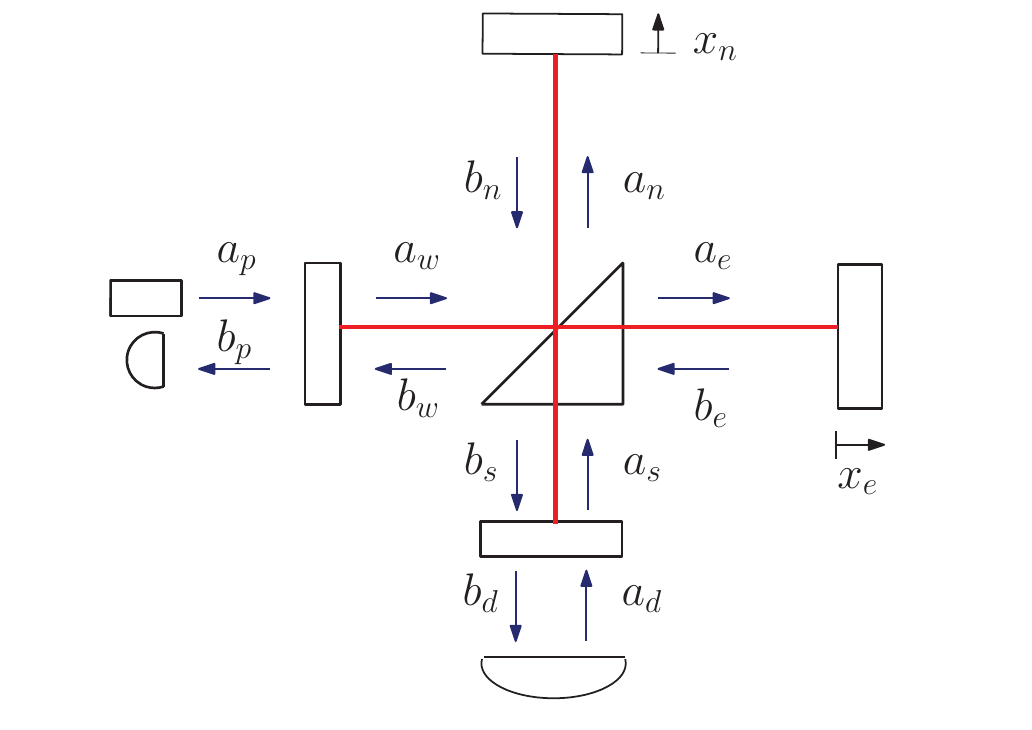}
		\caption{
		Michelson interferometer with power and signal recycling mirrors.
		}\label{Mich}
	\end{center}
\end{figure}

Let consider Michelson interferometer without FP cavities in arms but with power and signal recycling mirrors as shown on Fig.~\ref{Mich}. It can be easily  generalized on a case of aLIGO by redefining decay rates and detunings in this system.

The mirrors in east and north arms may move as free masses, whereas power and signal recycling mirror in west and south arms (with amplitude transmittances are $T_w,\, T_s$ correspondingly) are assumed to be unmovable. The interferometer is  pumped through west port. For simplicity we assume that mean distance $\ell$ between beam splitter and recycling mirrors in west and south arms is much smaller than mean distance $L$ between beam splitter and end mirrors in north and east arms: $\ell\ll L$. 

In case of complete balance optical paths in north and east arms are tuned so that whole output power returns through power recycling mirror in west arm and no average power goes through signal recycling mirror in south port.  In this case one can analyze symmetric and antisymmetric modes separately, in particular, symmetric mode interact with symmetric mechanical mode $(x_e +x_n)$ and antisymmetric one --- with $(x_e - x_n)$.  We analyze the non-balanced case when such separation is impossible.  

Below for complex amplitudes of fields we use notations on Fig.~\ref{Mich} denoting by capital letters mean amplitudes and by small letters --- small time dependent additions.

It is easy to obtain equations for mean amplitudes $A_w$ at power recycling mirror and $A_s$ at signal recycling one:
\begin{subequations}
\label{mean_PrSr2}
 \begin{align}
  A_w&\big(1- r_w\mathcal R_+\big) - A_s r_w\mathcal R_- = ie^{i\phi_w/2} T_w A_p ,     \\
  -A_w &r_s\mathcal R_-   + A_s\big(1-r_s\mathcal R_+\big) = i e^{i\phi_s/2}T_s A_d ,\\
  \mathcal R_+ &=\cos\delta\tau,\quad  \mathcal R_- =i \sin\delta\tau\,, \\
  \tau &= 2L/c, \quad r_{w,s}  = R_{w,s}e^{i\phi_{w,s}}
 \end{align}
\end{subequations}
Here $R_{w,s}$ are amplitude reflectivities of power and signal mirrors respectively, $\tau$ is round trip of light between beam splitters and end mirrors, $\delta$ is detuning introduced by displacements of north and east mirrors (in opposite directions), $\phi_{s,w}$ is round trip phase advance of wave traveling between beam splitter and power ($_w$) and signal ($_s$) recycling mirrors, $A_p$ is mean amplitude of pump laser, for generality we add term $\sim A_d$ describing possible pump through south port. 

By the same way one can obtain equations for small amplitudes in west and south arms in frequency domain
\begin{subequations}
  \label{aws}
 \begin{align}
    a_w &\left[1-r_w\mathcal R_+ e^{i\Omega\tau}\right] -a_s r_w \mathcal R_-e^{i\Omega\tau}=\\ 
    &\qquad = iT_w e^{i\phi_{w}/2}a_p  + r_w \, ikX_w, \nonumber\\
    - a_w & r_s \mathcal R_-e^{i\Omega\tau} + a_s\left[1-r_s\mathcal R_+ e^{i\Omega\tau}\right]=\\  
     &\qquad = iT_se^{i\phi_{s}/2} a_d  +  r_s \, ikX_s,\nonumber
 \end{align}
\end{subequations}
Here $\Omega$ is spectral frequency, due to strong unequality $\ell \ll L$ we assume that phases $\phi_{w,s}$ do not depend on  $\Omega$. $a_p,\ a_d$ describe zero point fluctuations of input field, $k$ is wave vector, values $X_{w,s}$ describe influence of displacements $x_e$ and $x_n$:

  \begin{subequations}
  \label{Xws}
  \begin{align}
  x_\pm & \equiv x_e\pm x_n,\\
  X_w &\equiv - e^{i\Omega\tau/2} \big(A_w\mathcal R_+ +A_s\mathcal R_-\big)x_+ -\\
      &\qquad - e^{i\Omega\tau/2} \big(A_w\mathcal R_-+A_s\mathcal R_+\big) x_- ,\quad	
	 \nonumber\\
  X_s &\equiv - e^{i\Omega\tau/2} (A_w\mathcal R_- +A_s\mathcal R_+)x_+ - \\
      &\qquad - e^{i\Omega\tau/2} (A_w\mathcal R_++A_s\mathcal R_-)x_-),\nonumber
  \end{align}
  \end{subequations}

In long wave approximation 
\begin{align}
\label{lwa}
\Omega\tau\ll 1,\quad \delta \tau \ll 1, \quad T_{w,s} \ll 1
\end{align}
we have $\mathcal R_+\simeq 1,\quad \mathcal R_-\simeq i\delta\tau$ and may simplify set \eqref{aws} as following
\begin{subequations}
  \label{aws2}
 \begin{align}
     \label{awEq}
    a_w & \left[\Gamma_w - i\Omega\right] -a_s \, i\delta = g_w,\ \\
      \label{asEq}
    - a_w & \, i\delta + a_s\left[\Gamma_s - i\Omega\right] = g_s,
  \end{align}
\end{subequations}
where

  \begin{align}
  \label{xwxsm}
    g_w & \equiv\frac{iT_w a_p - r_w \, ikX_w}{\tau r_w},\quad 
      g_s\equiv \frac{iT_s a_d - r_s \, ikX_s}{ \tau r_s}\nonumber\\
     X_w &= -A_w x_+ - A_s x_-,\quad
     X_s = -A_w x_- - A_s x_+.
  \end{align} 
  
One can write down the following approximate formulas for $\Gamma_w$ and $\Gamma_s$:
\begin{align}
 	\label{gammas}
 \Gamma_{w,s} & \simeq \frac{1 -R_{w,s} e^{i\phi_{w,s}}}{\tau R_{w,s} e^{i\phi_{w,s}}}=\gamma_{w,s} -i\delta_{w,s},\\
 	\gamma_{w,s} & \simeq \frac{1-R_{w,s} \cos\phi_{w,s}}{\tau} ,
	  \quad \delta_{w,s} \simeq \frac{\sin\phi_{w,s}}{\tau}\, .
\end{align}

 In case of zero detuning $\delta=0$ the set \eqref{aws2} transforms into equations of decoupled oscillators whereas non-zero $\delta$ introduces coupling.
 
 Important, the set \eqref{aws2} may be recalculated to equations for $e_\pm \rightarrow -(a_e \pm a_n)/\sqrt 2$ which are equivalent (with slightly different notations) to first four equations in set \eqref{setEV1}. Here we have to introduce symmetric and antisymmetric modes with sign ''minus'' because fields $a_{e,n}$ are defined near beam splitter whereas fields $e_{e,n}$ are defined near end mirrors of Fabry-Pero resonators.
 
 Now we can write down equations for ponderomotive forces acting on end mirrors of interferometer.  They can be expressed by next formula:
\begin{align}
\label{F_e}
&F_e = 2 \hbar k (A_e^{*}a_e + A_e a_e^{\dagger}), \\
\label{F_n}
&F_n = 2 \hbar k (A_n^{*}a_n + A_n a_n^{\dagger})\, .
\end{align}
For beam splitter we can use following relations (similar to (\ref{eil_PM_0}, \ref{Defs})):

\begin{align}
\label{relA}
&A_e = -\frac{A_w + A_s}{\sqrt{2}}, \quad A_n = -\frac{A_w - A_s}{\sqrt{2}}, \\
&a_e = -\frac{a_w + a_s}{\sqrt{2}}, \quad a_n =- \frac{ a_w - a_s}{\sqrt{2}}\, .
\end{align}

And we can write:
\begin{align}
\label{F-}
F_- =& \frac{F_e - F_n}{2} = \\ \nonumber & = \hbar k (A_w^* a_s + A_s^*a_w + A_w a_s^{\dagger} + A_s a_w^{\dagger}) \, .
\end{align}
Equation of motion for antisymmetric mode can be expressed in next form:
\begin{align}
\label{mot}
\hbar k (A_w^* a_s + A_s^*a_w + A_w a_s^{\dagger} + A_s a_w^{\dagger}) + \mu \Omega^2 z_- (\Omega) = 0.
\end{align}
This equation is equivalent to \eqref{mot1} with corresponding substitutions mentioned above.

\subsection{Michelson-Sagnac interferometer}

Let now consider Michelson-Sagnac interferometer with power and signal recycling mirrors as presented on a Fig.\ref{ms}. Similarly one can obtain a set of equations for small amplitudes in long wave approximation:
\begin{subequations}
    \label{char1lV}
 \begin{align}
   [\Gamma_w - i \Omega]\,a_w - id\, a_s &= g_w, \\ 
- id^*\, a_w  + [\Gamma_s - i\Omega]\, a_s &= g_s,
 \end{align}
\end{subequations}
where

\begin{align}
  &\Gamma_w = \frac{1 - r_w(iT_z + \tilde R_+ R_z)}{ r_w\tau'(iT_z + R_z)}\simeq 
	\frac{1 - \tilde r_w}{ \tilde r_w\tau'},\\ & \tilde r_w \equiv r_w(R_z +iT_z ),\quad \tilde R_+ = \cos  \delta \tau^{'} \to 1,\\
  & \Gamma_s = \frac{1 - r_s(-iT_z + \tilde R_+ R_z)}{ r_s\tau'(-iT_z +  R_z)}\simeq
	\frac{1 - \tilde r_s}{ \tilde r_s\tau'},\\& \tilde r_s \equiv r_s( R_z -iT_z ),
   \\
  &  d \equiv  \frac{ \delta \,R_z}{iT_z + R_z}, \quad
  \label{gw}
    g_w = \frac{i T_w a_p + r_w ik X_w}{\tau' r_w (R_z+iT_z)},\\
    &g_s = \frac{i T_s a_d + r_s ikX_s}{\tau' r_s (R_z-iT_z)} \, . 
    \label{gs}   
\end{align}

In \eqref{gw},\eqref{gs} values $X_{w,s}$ describe influence of displacement $x$ of membrane.
\begin{subequations}
     \label{XwsL}
    \begin{align}
    X_w = 2 A_s R_z x, \\ 
    X_s =  2 A_w R_z x \, .
    \end{align} 
\end{subequations}
We introduced $\tau^{'} = \frac{2(L+l)}{c}$. Here $L$ --- distance between beamsplitter and east (or north) mirror, $l$ --- distance between east (or north) mirror and membrane, $R_z$  --- amplitude reflectivity of membrane and $T_z = \sqrt{1-R_z^2}$ --- its amplitude transparency.

Formally, $\gamma_w,\ \gamma_s$ are complex values, however, their imaginary parts are much smaller than real ones --- due to
condition $\delta\tau'\ll 1$.  In long waves approximations we may put $\tilde R_+ \simeq 1$.

Now we have to write equations for ponderomotive forces acting on membrane:
\begin{align}
\label{F_a}
 & F_a = \hbar k (A_e^* a_e + A_e a_e^{\dagger} - A_n^* a_n - A_n a_n^{\dagger}), \\
 \label{F_b}
 & F_b = \hbar k (B_e^* b_e + B_e b_e^{\dagger} - B_n^* b_n - B_n b_n^{\dagger}) \, .
\end{align}

Using following expressions in long wavelenght approximation:

\begin{align}
&b_w = a_w \big[  i T_z + R_z \big] + A_s R_z i 2 k x, \\
&b_s = a_s \big[- i T_z + R_z] + A_w R_z 2 i k x, \\
&B_w = A_w \big[ i T_z + R_z \big], \\
&B_s = A_s \big[ -i T_z + R_z \big],
\end{align}
and relations for beam splitter similar to (\ref{eil_PM_0}, \ref{Defs})
\begin{subequations}
\begin{align}
&A_e = -\frac{A_w + A_s}{\sqrt{2}}, \quad A_n = -\frac{A_w - A_s}{\sqrt{2}}, \\
&a_e = - \frac{a_w + a_s}{\sqrt{2}}, \quad a_n = - \frac{a_w - a_s}{\sqrt{2}}, \\
&B_e = -\frac{B_w + B_s}{\sqrt{2}}, \quad B_n = -\frac{B_w - B_s}{\sqrt{2}}, \\
&b_e = - \frac{b_w + b_s}{\sqrt{2}}, \quad b_n = - \frac{b_w - b_s}{\sqrt{2}}, 
\end{align}
\end{subequations}
we can rewrite equations \eqref{F_a} and \eqref{F_b} in next form:
\begin{align}
&F_a = \hbar k \big( A_w^* a_s + A_w a_s^{\dagger} + A_s^* a_w + A_s a_w^{\dagger} \big), \\ \nonumber
&F_b = \hbar k \big( B_w^* b_s + B_w b_s^{\dagger} + B_s^* b_w + B_s b_w^{\dagger} \big) = \\ \nonumber
& = \hbar k \big(A_w^*a_s[-iT_z+R_z]^2 + A_wa_s^\dag[iT_z+R_z]^2 + \\ &+ A_s^*a_w[iT_z+R_z]^2 + A_s a_w^\dag [-iT_z+R_z]^2\big)
\end{align}
And the total force acting on membrane can be expressed by next formula:
\begin{align}
&F  \equiv F_a+F_b = \\ \nonumber &=  2\hbar k\,R_z \big(A_w^*a_s[-iT_z+R_z] + A_wa_s^\dag[iT_z+R_z] + \\ \nonumber  &+ A_s^*a_w[iT_z+R_z] + A_s a_w^\dag [-iT_z+R_z]\big). 
\end{align}

Now we can write down equation of motion for membrane:
\begin{align}
F + m \Omega^2 x (\Omega) = 0.
\end{align}
So we may state that formulas \eqref{char1lV} for Michelson-Sagnac interferometer (MSI) are equivalent to formulas \eqref{aws2} for antisymmetric mode of Michelson interferometer (DMMI):
\begin{itemize}
 \item Formulas for MSI transform for DMMI in limit $R_z\to 1$.
 \item Formulas for DMMI transforms into formulas for MSI with following substitutions in definitions of $\gamma_{w,s}$ and 
effective detuning $d$

 \begin{align}
  &r_{w,s} \to  r_{w,s}\big(R_z-iT_z\big) ,\quad \tilde R_+ \to 1,\\
    &\delta \to d \equiv \alpha\,\delta,\quad \alpha \equiv  \frac{ R_z}{iT_z + R_z} \, .
 \end{align}

  \item Formulas for DMMI transforms into formulas for MSI with substitutions in definitions of right parts 
  $g_{w,s}$ according to \eqref{gw}, \eqref{gs} and \eqref{Xws}.
 \end{itemize}


\bibliography{modes}

\end{document}